\documentclass[prd,aps,preprint,tightenlines,superscriptaddress]{revtex4}

\usepackage{amsmath}
\usepackage{epsfig}
\usepackage{axodraw}

\newcommand{\insertfig}[2]{\mbox{\epsfxsize=#1cm \epsfbox{#2.eps}}}
\preprint{DOE/ER/40762-352}

\begin{document}

\title{An SO(10) GUT Model With Lopsided Mass Matrix \\
and Neutrino Mixing Angle $\theta_{13}$}
\author{Xiangdong Ji}
\affiliation{Department of Physics, University of Maryland, College Park, Maryland 20742}
\author{Yingchuan Li}
\affiliation{Department of Physics, University of Maryland, College Park, Maryland 20742}
\author{R. N. Mohapatra}
\affiliation{Department of Physics, University of Maryland,
College Park, Maryland 20742}
\date{\today}

\begin{abstract}
An alternative supersymmetric SO(10) grand unification model with
lopsided fermion mass matrices is introduced. It generates a large
solar-neutrino-mixing angle through the neutrinos' Dirac mass
matrix constrained by the SO(10) group structure, avoiding the
fine-tuning required in the Majorana mass matrix of right-handed
neutrinos. The model fits well the known data on masses and
mixings of quarks and leptons, and predicts a sizable lepton
mixing $\sin^22\theta_{13}\simeq 0.074$, which is significantly
larger than that of the original lopsided model.
\end{abstract}

\maketitle

The discovery of neutrino oscillation has opened up a fascinating
window for physics beyond the standard model. Experimental data on
neutrino mass differences and mixings help to constrain various
theoretical models of new physics. Assuming three light flavors,
the lepton Pontecorvo-Maki-Nakagawa-Sakata (PMNS) mixing matrix is
characterized by three mixing angles $(\theta_{12},\theta_{13},
\theta_{23})$ and three CP-violating phases when neutrinos are
Majorana fermions. The atmospheric and accelerator neutrino data
have determined $\theta_{23}$ to a good accuracy, and the solar
neutrino and reactor neutrino experiments have measured
$\theta_{12}$ with an even better precision
~\cite{McKeown:2004yq}. These results have already helped to
eliminate a large class of neutrino mass matrix models in the
literature. The CHOOZ reactor experiment has found that
$\rm{sin}^22\theta_{13}$, if non-zero, should be smaller than
$0.1$ \cite{Apollonio:2002gd}. The next generation of neutrino
experiments under proposal aims to push the limit to $\sin^22
\theta_{13}\sim 0.01$ \cite{Anderson:2004pk,Heeger:2003aa}, which
undoubtedly will teach us a great deal about the mechanism of
neutrino mass generation.

If small neutrino masses are assumed to arise from the seesaw
mechanism \cite{seesaw}, the first thing one learns from present
data is that the seesaw scale (the mass of the right-handed
neutrinos) must be very high (close to $10^{15}$ GeV or so). This
strongly suggests that the seesaw scale may be connected  with one
of the leading ideas for new physics beyond the standard model,
i.e., supersymmetric grand unification theory
 (GUT) according to which all forces and matter unify at very short
distances corresponding to energies of order $10^{16}$ GeV. Since
the GUT models unify the quarks and leptons they build in more
constraints and have better predictive power
\cite{Mohapatra:2004za} which can connect neutrino parameters to
the well-determined quark parameters.

The most minimal GUT models that incorporate the seesaw mechanism
are based on SUSY SO(10) since its {\bf 16} dimensional spinor
representation contains all fermions of the standard model along
with the right-handed neutrino needed for this purpose as well as
the fact that it has $B-L$ as a subgroup whose breaking gives rise
to masses to the right-handed neutrinos. Depending upon which set
of Higgs multiplets is chosen to break the $B-L$ subgroup of
SO(10), there emerge two classes of SO(10) models: one that uses
${\bf 10}_H$, ${\bf 126}_H$, ${\bf \overline{126}}_H$ and ${\bf
120}_H$ \cite{Chen:2000fp,Goh:2003sy}, and the other that uses
${\bf 10}_H$, ${\bf 16}_H$, ${\bf \overline{16}}_H$ and ${\bf
45}_H$ \cite{Babu:1998wi,Albright:1998vf,Blazek:1999hz}. While
most of these models are quite successful in fitting and
predicting the known experimental masses and mixing angles of
leptons and quarks, they predict very different values for the
poorly-known neutrino mixing angle $\theta_{13}$---majority of
models with high-dimensional Higgses tend to yield $\theta_{13}$
close to the current experimental upper bound and majority of
those with low-dimensional Higgses generally result in a small
$\theta_{13}$, hence, a small CP violation in the lepton sector.
Thus it appears that $\theta_{13}$ might be an excellent
observable to differentiate the two classes of SO(10) models.

Consider, for example, the SO(10) model with low-dimensional
Higgses and the so-called lopsided fermion mass matrices proposed
by Albright, Babu, and Barr
\cite{Albright:1998vf,Albright:2001LMA,Albright:2000dk}. The
lopsidedness built within the Yukawa couplings between the second
and third families generates, among other interesting physical
consequences, the large atmospheric-neutrino mixing angle
$\theta_{23}$ while keeping $V_{cb}$ in the
Cabbibo-Kobayashi-Moskawa (CKM) matrix small. The large
solar-neutrino mixing angle $\theta_{12}$, however, is generated
less elegantly. It is obtained through a fine-tuning which not
only requires the $2$-$3$ and $3$-$2$ entries in the Majorana mass
matrix $M_R$ of the right-handed neutrinos to be of order of
parameter $\epsilon$ appearing in Dirac mass matrices of quarks
and leptons, but also requires them to be exactly $-\epsilon$
\cite{Albright:2001LMA}. By varying the four parameters in the
$M_R$ \cite{Albright:2000dk}, the predication of $\theta_{13}$
from this model was found to lie in the range of $10^{-5}\leq
\sin^2\theta_{13}\leq10^{-2}$. A narrower range of $0.002\leq
\sin^2\theta_{13}\leq0.003$ is obtained when constraints are
imposed on the parameter space.  If $\bar{\nu}_e$ disappearance is
observed in the next generation of short baseline reactor
experiments \cite{Heeger:2003aa}, the original lopsided model
would be ruled out.

Given that the lopsided fermion matrix model is one of the most
successful GUT theories incorporating all the known experimental
facts, two obvious questions arise immediately. First, is there a
more natural way to realize the large solar-neutrino mixing angle
without fine tuning? And second, if such an alternative model
exists, is $\theta_{13}$ consistently small? In this paper, we
present a modified lopsided model which uses an alternative
mechanism to generate the solar-neutrino mixing angle. We assume
that the right-handed neutrino Majorana mass matrix $M_R$ has a
simple diagonal structure, and introduce additional off-diagonal
couplings in the upper-type-quark and neutrino Dirac mass matrices
to generate $1$-$2$ rotation. We found that all the fermion masses
and mixing angles can be fitted well in the new model. The mixing
angle $\theta_{13}$, however, is close to the upper limit from the
CHOOZ experiment and therefore definitely within the reach of next
generation reactor experiments.

Before we present our model for fermion mass matrices, it is
instructive to review some of the salient features of the SUSY
SO(10) model with lopsided fermion mass matrices
\cite{Albright:1998vf,Albright:2001LMA}. Through couplings with a
set of Higgs multiplets ${\bf 10}_H$, ${\bf 16}_H$, ${\bf
\overline{16}}_H$ and ${\bf 45}_H$ and the constraint from the
flavor $U(1)\times Z_2\times Z_2$ symmetry, the fermion mass
matrices have the following forms,
\begin{eqnarray} U&=&\begin{pmatrix} \eta & 0 & 0
\\ 0 & 0 & \epsilon/3 \\ 0 & -\epsilon/3 & 1
\end{pmatrix}M_U \ ,
~~~~~~~~~~~ N=\begin{pmatrix} \eta & 0 & 0 \\ 0 & 0 & -\epsilon \\
0 &
\epsilon & 1 \end{pmatrix}M_U \ , \nonumber \\
D&=&\begin{pmatrix} \eta & \delta & \delta'e^{i\phi} \\
\delta & 0 & \sigma+\epsilon/3 \\ \delta'e^{i\phi} & -\epsilon/3 &
1
\end{pmatrix}M_D \ , ~~~~
L=\begin{pmatrix} \eta & \delta & \delta'e^{i\phi} \\
\delta & 0 & -\epsilon \\ \delta'e^{i\phi} & \sigma+\epsilon & 1
\end{pmatrix}M_D \ ,~~ \nonumber \\
M_R&=&\begin{pmatrix} c^2\eta^2 & -b\epsilon\eta & a\eta \\
-b\epsilon\eta & \epsilon^2 & -\epsilon \\ a\eta & -\epsilon & 1
\end{pmatrix}\Lambda_R \ ,
\end{eqnarray}
where $U$, $D$, $L$, and $N$ denote up-type-quark,
down-type-quark, charged lepton, and neutrino Dirac mass matrices,
respectively, and $M_R$ is the Majorana mass matrix of the
right-handed neutrinos. As explained in
\cite{Albright:1998vf,Albright:2001LMA}, the various entries in
the mass matrices come from different SO(10) invariants in the
superpotential, e.g., $\eta$ from ${\bf 16}_1{\bf 16}_1 {\bf
10}_H$; $\epsilon$ from ${\bf 16}_2{\bf 16}_3 {\bf 10}_H{\bf
45}_H$, $\delta,\delta'$ from ${\bf 16}_1{\bf 16}_{2,3} {\bf
16}_H{\bf 16'}_H$; and $\sigma$ from ${\bf 16}_2{\bf 16}_H {\bf
16}_3{\bf 16'}_H$.

The parameter $\sigma$ is of order one, signaling the lopsidedness
between the second and third families in $D$ and $L$. This feature
leads to a large left-handed neutrino mixing in the PMNS matrix
and a small left-handed quark mixing shown in the CKM matrix. The
parameter $\epsilon$ is one order-of-magnitude smaller than
$\sigma$ and generates the hierarchy between the second and third
families. In extending to the first family, $\delta$ and $\delta'$
were introduced into the $D$ and $L$. The large solar-neutrino
mixing angle is from the left-handed neutrino seesaw mass matrix
which in turn depends on a very specific structure in $M_R$.

Since the lepton mixing matrix is defined as
\begin{equation}
    U_{PNMS} = U^\dagger_L U_\nu \ ,
\end{equation}
the large solar mixing angle can either be generated from
$U_L^\dagger$ or $U_\nu$ or a combination of both. If there is a
non-vanishing $1$-$2$ rotation from $U_\nu$, it can either be
generated from the Dirac mass matrix of the left-handed neutrinos
or from the Majorana mass matrix of the right-handed neutrinos or
a combination of both. In the following, we focus on the
possibilities in which one of the matrices generates a large
solar-neutrino mixing angle, keeping in mind though that a general
situation might involve a mixture of the extreme cases. In the
fermion mass model in Eq. (1), the large solar-neutrino mixing is
induced mainly by the right-handed neutrino mass matrix.

Thus, an alternative possibility is to produce the large
solar-neutrino mixing from the charged lepton matrix. In fact, in
Ref. \cite{Babu:2001cv}, a model was proposed in which both large
solar and atmospheric neutrino mixings are generated from the
lopsided charged-lepton mass matrix. The value of
$\sin^22\theta_{13}$ is again found to be small, 0.01 or less.

Here we study yet a third possibility of generating a large size
1-2 rotation in the lepton mixing from the neutrinos' Dirac mass
matrix $N$. The easiest way to achieve this might be to use a
lopsided structure in the 1-2 entries of $N$. However, this is
impossible in group theory of SO(10). A large rotation, however,
can be generated through 1-3 and 2-3 entries without affecting,
for example, the quark mass hierarchy between the first and second
generations. Thus we introduce the following modifications of the
up-type quark and neutrino mass matrices in Eq. (1),
\begin{eqnarray}
U&=&\begin{pmatrix} \eta & 0 & \kappa+\rho/3 \\ 0 & 0 & \omega \\
\kappa-\rho/3 & \omega & 1 \end{pmatrix}M_U,~~~~
N=\begin{pmatrix} \eta & 0 & \kappa-\rho \\ 0 & 0 & \omega \\
\kappa+\rho & \omega & 1 \end{pmatrix}M_U, \nonumber \\
 M_R&=&\begin{pmatrix} a & 0 & 0 \\
0 & b & 0 \\ 0 & 0 & 1
\end{pmatrix}\Lambda_R,
\end{eqnarray}
The symmetric entries $\omega$ and $\kappa$ in $U$ and $N$ can be
generated from the dimension-5 operator ${\bf 16}_i{\bf 16}_j[{\bf
\overline{16}}_H{\bf \overline{16}}'_H]_{10}$, and the
antisymmetric $\rho$ entries in $U$ and $N$ are from dimension-6
operator ${\bf 16}_i{\bf 16}_j[{\bf \overline{16}}_H{\bf
\overline{16}}'_H]_{10}{\bf 45}_H$, where the subscript $10$
indicate that the spinor Higgses are coupled to 10 of SO(10).
Because of the modification, the $\epsilon$ entries in $D$ and $L$
now must be generated from dimension-6 operator ${\bf 16}_i{\bf
16}_j[{\bf 16}_H{\bf 16}'_H]_{10}{\bf 45}_H$. We assume as in the
past that ${\bf 45}_H$ Higgs develops a vacuum expectation value
(VEV) in the $B-L$ direction.  ${\bf 16}_H$ and ${\bf
\overline{16}}_H$ are the Higgs spinors which break the SO(10) to
SU(5) by taking the VEV in the singlet direction of SU(5). The
second pair of ${\bf 16}'_H$ and ${\bf \overline{16}}'_H$ develop
VEV in ${\bf \overline{5}}$ and ${\bf 5}$ of SU(5), respectively,
and therefore the operators involving ${\bf \overline{16}}'_H$ and
${\bf 16}'_H$ contribute to up and down sectors as weak doublets,
respectively.

Usually a rotation is connected with the mass spectrum. However,
in our case the $1$-$2$ rotation angle from $U$ will be combined
with the $1$-$2$ rotation from $D$ to obtain the Cabibbo angle
$\theta^{c}$, and a constraint from the up-type quark spectrum
must be avoided. Thus, the first two families in the $U$ and $N$
cannot be coupled to each other directly, but can be coupled
indirectly through the third family. The $1$-$2$ rotations in $U$
and $N$ generated from this way are proportional to the ratios
$\gamma\equiv(\kappa-\rho/3)/\omega$ and
$\gamma'\equiv(\kappa+\rho)/\omega$, respectively.

Taking the approximation $\eta=0$, the dependence of various mass
ratios and CKM elements on parameters can be seen roughly from the
following approximate expressions (the superscript 0 indicates the
relevant quantity is at GUT scale)
\begin{eqnarray}
m^0_b/m^0_{\tau}&\simeq& 1-
\frac{2}{3}\frac{\sigma}{\sigma^2+1}\epsilon \ , ~~ \nonumber \\
m^0_u/m^0_{t}&\simeq& 0 \ , ~~ \nonumber \\
m^0_c/m^0_{t}&\simeq& (1+\gamma^2)\omega^2\ , ~~ \nonumber \\
m^0_{\mu}/m^0_{\tau}&\simeq&
\epsilon\frac{\sigma}{\sigma^2+1} \ ,~~\nonumber\\
m^0_s/m^0_b&\simeq& \frac{1}{3}\epsilon\frac{\sigma}{\sigma^2+1} \
,
\nonumber \\
 m^0_e/m^0_{\mu}&\simeq& \frac{1}{9}t_Lt_R \ , \nonumber \\
  m^0_d/m^0_s
&\simeq& t_Lt_R, \nonumber \\
V^0_{cb}&\simeq& -\sqrt{1+\gamma^2}\omega -
\frac{1}{\sqrt{1+\gamma^2}}\frac{\epsilon}{3(1+\sigma^2)} \ ,
\nonumber \\
V^0_{us}&\simeq& \frac{1}{\sqrt{1+\gamma^2}}(-\gamma+t_L e^{i
\theta}) \ , ~~\nonumber \\
V^0_{ub}&\simeq&
\frac{1}{\sqrt{1+\gamma^2}}\frac{\epsilon}{3(\sigma^2+1)}(\gamma-t_L
e^{i \theta}+\sqrt{1+\sigma^2}t_R) \ ,
\end{eqnarray}
where $t_L$, $t_R$ and $\theta$ are defined as
$t_Le^{i\theta}\equiv
3(\delta-\sigma\delta'e^{i\phi})/(\sigma\epsilon)$ and $t_R\equiv
3\delta\sqrt{\sigma^2+1}/(\sigma\epsilon)$. The expressions for
mass ratios in down-type quark and charged lepton sectors are the
same as those in the original lopsided model. The expressions for
$m^0_c/m^0_t$ and elements in CKM matrix are new. These
approximations allow us to design strategies to fit various
parameters to experimental data.

First, we use the up-type quark and lepton spectra and the
parameters in the CKM matrix to determine 10 parameters $\sigma$,
$\epsilon$, $\delta$, $\delta'$, $\phi$, $\omega$, $\gamma$,
$\eta$, $M_U$ and $M_D$. Our best fit yields $\sigma$ and
$\epsilon$ approximately the same as those in the original
lopsided model, and thus the successful prediction for the mass
ratios $m^0_{\mu}/m^0_{\tau}$ and $m^0_s/m^0_b$ are kept. The two
CKM elements $|V^0_{us}|$ and $|V^0_{ub}|$, together with the CP
violation phase $\delta_{CP}$ and the constraint on the product
$t_Lt_R$ from mass ratio $m^0_{e}/m^0_{\mu}$, can fix the $t_L$,
$t_R$, $\gamma$ and $\theta$. Then $\omega$ and $\eta$ can be
fixed from $m^0_c/m^0_t$ and $m^0_u$, respectively. The down-type
quark mass spectrum come out as predictions.

To see the dependence of the lepton mixing PMNS matrix on various
parameters, we construct the Majorana mass matrix of left-handed
neutrino from the see-saw mechanism\cite{seesaw}, $m_{\nu}=-NM^{-1}_RN$,
\begin{eqnarray}
m_{\nu}=-\begin{pmatrix} \eta^2/a+(\kappa+\rho)^2 &
(\kappa+\rho)\omega & \eta(\kappa-\rho)/a+(\kappa+\rho)
\\ (\kappa+\rho)\omega & \omega^2 & \omega
\\
\eta(\kappa-\rho)/a+(\kappa+\rho) & \omega & 1+
(\kappa-\rho)^2/a+\omega^2/b\end{pmatrix}M^2_U/\Lambda_R \ ,
\end{eqnarray}
which depends on the four unknown parameters, $\gamma'$,
$\Lambda_R$, $a$ and $b$. With parameter $a$ taking a reasonably
large value, say, order of 0.001 or larger, the $\eta$ dependent
terms can be neglected. Then one readily sees that the $m_{\nu}$
matrix can be diagonalized by a 1-2 rotation of angle
$\theta^{\nu}_{12}$ with $\rm{tan} ~\theta^{\nu}_{12}=\gamma'$,
and followed by a 2-3 rotation by angle $\theta^{\nu}_{23}$, with
\begin{equation}
\tan
2\theta^{\nu}_{23}=\frac{2\sqrt{1+\gamma'^2}\omega}{1+(\kappa-\rho)^2/a+\omega^2/b-(1+\gamma'^2)\omega^2}
\ .
\end{equation}
The neutrino Majorana masses of the second and the third families
are
\begin{eqnarray}
m_{\nu2}&=&-\left[(1+\gamma'^2)\omega^2+\sqrt{1+\gamma'^2}\omega
\left(\rm{cot}2\theta^{\nu}_{23} - \rm{csc}2\theta^{\nu}_{23}
\right)\right] M_U^2/\Lambda_R \nonumber \ ,
\\
m_{\nu3}&=&-\left[(1+\gamma'^2)\omega^2+\sqrt{1+\gamma'^2}\omega
\left(\rm{cot}2\theta^{\nu}_{23} + \rm{csc}2\theta^{\nu}_{23}
\right)\right]M_U^2/\Lambda_R \ ,
\end{eqnarray}
with $m_{\nu1}=0$ as the result of the approximation $\eta=0$.
Therefore, the present model constrains the neutrino mass spectrum
as hierarchial, which means that the parameters in the
light-neutrino mass matrix, the mass eigenvalues and mixings, do
not run significantly from GUT to low-energy scales. The mass
difference $\Delta m^2_{\nu 12}$ can be used to fix the
right-handed neutrino mass scale $\Lambda_R$.

Taking into account rotations from matrices $m_{\nu}$ and $L$,  we
arrive at the elements in the PMNS matrix,
\begin{eqnarray}
U_{e2}&=&\left(\frac{\gamma'}{\sqrt{1+\gamma'^2}}-\frac{t_R}{3}
\frac{1}{\sqrt{1+\gamma'^2}}\frac{1}{\sqrt{1+\sigma^2}}
\right)\cos\theta^{\nu}_{23}-\frac{t_R}{3}
\frac{\sigma}{\sqrt{1+\sigma^2}}\rm{sin}\theta^{\nu}_{23} \ ,
\nonumber \\
U_{\mu3}&=&-\frac{\sigma}{\sqrt{1+\sigma^2}}\cos\theta^{\nu}_{23}
+\frac{1}{\sqrt{1+\gamma'^2}}\left(\gamma'\frac{t_R}{3}+
\frac{1}{\sqrt{1+\sigma^2}}
 \right)\rm{sin}\theta^{\nu}_{23} \ , \nonumber \\
U_{e3}&=&\frac{t_R}{3}\frac{\sigma}{\sqrt{1+\sigma^2}}\cos\theta^{\nu}_{23}
+\left(\frac{\gamma'}{\sqrt{1+\gamma'^2}} -
\frac{t_R}{3}\frac{1}{\sqrt{1+\sigma^2}\sqrt{1+\gamma'^2}}\right)
\rm{sin}\theta^{\nu}_{23} \ .
\end{eqnarray}
The data on the solar-neutrino mixing $U_{e2}$, together with the
ratio of mass differences, $\Delta m^2_{\nu12}/\Delta
m^2_{\nu23}=m^2_{\nu2}/(m^2_{\nu3}-m^2_{\nu2})$, can fix $\gamma'$
and $\theta^{\nu}_{23}$, where the latter depends on a combination
of $a$ and $b$. Having fixed $\gamma'$ and parameters in $M_R$,
the atmospheric-neutrino mixing $U_{\mu3}$ and $U_{e3}$ are
obtained as predictions.

We summarize our input and detailed fits as follows. For CKM
matrix elements, we take $|V_{us}|=0.224$, $|V_{ub}|=0.0037$,
$|V_{cb}|=0.042$, and $\delta_{CP}=60^\circ$ as inputs at
electro-weak scale. With a running factor of $0.8853$ for
$|V_{ub}|$, and $|V_{cb}|$ taken into account, we have
$|V^0_{ub}|=0.0033$ and $|V^0_{cb}|=0.037$ at GUT scale. For
charged lepton masses and up-type quark masses, we take the values
at GUT scale corresponding to $\rm{tan}\beta=10$ from Ref.
\cite{das}. For neutrino oscillation data, we take the
solar-neutrino angle to be $\theta_{\rm solar}=32.5^\circ$ and
mass square differences as $\Delta
m^2_{\nu12}=7.9\times10^{-5}\rm{eV}^2$ and $\Delta
m^2_{\nu23}=2.4\times10^{-3}\rm{eV}^2$. The result for the 12
fitted parameters is
\begin{eqnarray} \sigma&=&1.83, ~~\epsilon=0.1446, ~~
 \delta=0.01, \nonumber \\
 \delta'&=&0.014,~~\phi=27.9^\circ, ~~\eta=1.02\times10^{-5}, \nonumber \\
 \omega&=&-0.0466,~~\rho=0.0092, ~~\kappa=0.0191, \nonumber \\
 M_{U}&=&82.2~ {\rm GeV},~~ M_{D}=583.5~{\rm MeV}, ~~\Lambda_R=1.85\times10^{13}~{\rm GeV}
\end{eqnarray}
There is a combined constraint on $a$ and $b$, and thus the
right-handed Majorana mass spectrum is not well determined. As
examples, if $a=b$, $a=-2.039\times 10^{-3}$; and if $a=1$,
$b=-1.951\times 10^{-3}$.

We show the result for the down-type quark masses and right-handed
Majorana neutrino masses (taking $a=b$) as follows,
\begin{eqnarray}
&& m^0_d=1.08~{\rm MeV}, ~~ m^0_s=25.97~{\rm MeV}, ~~ m^0_b=1.242~
{\rm GeV}, \nonumber \\
&&M_1=3.77\times10^{10}{\rm GeV}, ~~ M_2=3.77\times10^{10}{\rm
GeV}, ~~ M_3=1.85\times10^{13}{\rm GeV} \ .
\end{eqnarray}
The predictions for the mixing angles in the PMNS matrix are,
\begin{equation}
\sin^2\theta_{\rm atm}=0.49, ~~ \sin^22\theta_{13}=0.074 \ .
\end{equation}
The result for $\theta_{\rm atm}$ is particularly interesting:
Although the lopsided mass matrix model is built to generate a
large atmospheric-neutrino mixing angle, the charged lepton mass
matrix alone produces a 2-3 rotation of $63^{\circ}$ instead of
$45^{\circ}$ because of the constraint from the lepton mass
spectrum. With an additional rotation $\theta^{\nu}_{23} \simeq
21^{\circ}$ fixed mainly from the ratio of mass differences
$\Delta m^2_{\nu12}/\Delta m^2_{\nu23}$, the nearly maximal
atmospheric mixing $44.6^{\circ}$ comes out as a prediction. If
one releases the best-fit value of $\Delta m^2_{\nu12}$ and
$\Delta m^2_{\nu23}$ and only imposes the $3\sigma$ constraint as
$7.1\times10^{-5}{\rm eV}^2\leq\Delta
m^2_{\nu12}\leq8.9\times10^{-5}{\rm eV}^2$ and
$1.4\times10^{-3}{\rm eV}^2\leq\Delta
m^2_{\nu23}\leq3.3\times10^{-3}{\rm eV}^2$, one would obtain, as
shown in Fig. 1, $0.44\leq\sin^2\theta_{\rm atm}\leq0.52$ which is
well within the $1\sigma$ limit, and
$0.055\leq\sin^22\theta_{13}\leq0.110$ which, as a whole region,
lies in the scope of next generation of reactor experiments.
\begin{figure}[t]
\begin{center}
\mbox{
\begin{picture}(0,130)(60,0)

\put(-140,15){\insertfig{6.5}{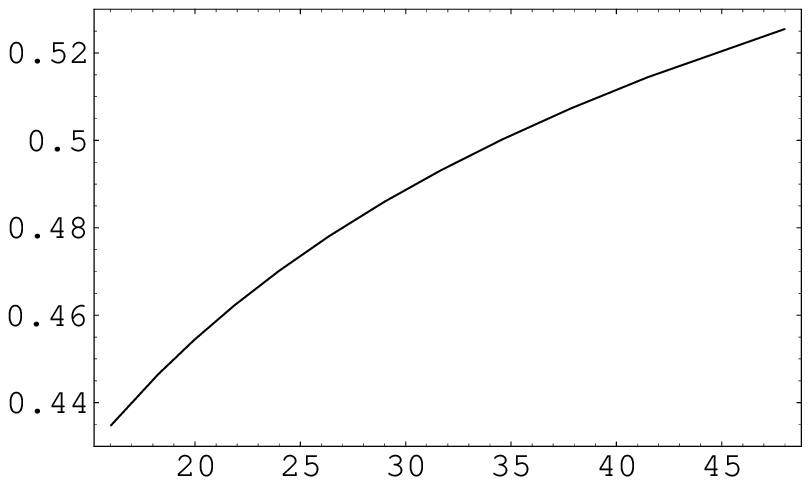}}

\put(100,15){\insertfig{6.5}{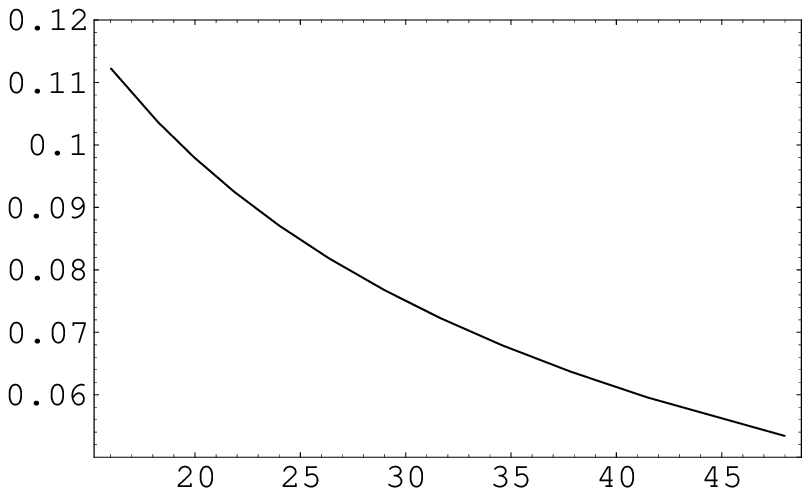}}

\Text(-160,110)[c]{$\sin^2 \theta_{\rm atm}$}
\Text(80,110)[c]{$\sin^2 2\theta_{13}$}

\Text(-40,2)[c]{$\Delta m^2_{\nu23}/\Delta m^2_{\nu12}$}
\Text(200,2)[c]{$\Delta m^2_{\nu23}/\Delta m^2_{\nu12}$}

\end{picture}
}
\end{center}
\caption{The predictions of $\sin^2 \theta_{\rm atm}$ and $\sin^2
2\theta_{13}$ against the mass square difference ratio $\Delta
m^2_{\nu23}/\Delta m^2_{\nu12}$. The region of $\Delta
m^2_{\nu23}/\Delta m^2_{\nu12}$ is obtained from the values of
$\Delta m^2_{\nu23}$ and $\Delta m^2_{\nu12}$ within their
3$\sigma$ limits.}
\end{figure}

Finally, we make some remarks on CP phases in the lepton mixing
matrix in our model. Since we essentially treat all our parameters
as real, the CP-violation in the lepton sector is essentially
absent. One might wonder why the $\phi$ phase in $L$ and $D$,
which generates the CP phase in the CKM matrix, does not give
contribution to the CP phases in the PMNS matrix. The answer is
the lopsided structure of the $L$ matrix. In fact, the unitary
matrix diagonalizing $L^\dagger L$ is nearly real, whereas that
diagonalizing $D^\dagger D$ is not. There is, of course, some
trivial CP phases due to specific choices of flavor basis. For
example, we find $\delta_{CP}^{PMNS}\sim \pi$, and some of the
Majorana phases are close to $\pi/2$, all of which are believed to
be artifacts of the model.

In summary, we have presented an SUSY SO(10) GUT model for the
fermion masses and mixings, which is developed from the original
lopsided model of Albright, Babu and Barr \cite{Albright:1998vf}.
It contains 13 parameters. After fitting them to experimental
data, it yields a number of predictions. Whenever the experimental
data are available, they work well. Most interestingly, the
model predicts a $\sin^22\theta_{13}$ around 0.074, which is
significantly larger than that from any of previous lopsided
models. It can surely be tested through the next generation of
reactor neutrino experiments. It will also have its characteristic
predictions for lepton flavor violation, leptogenesis as well as proton
decay. These issues are currently under investigation.

X. Ji and Y. Li are partially supported by the U. S. Department of
Energy via grant DE-FG02-93ER-40762 and by National Natural
Science Foundation of China (NSFC). R. Mohapatra is supported by
National Science Foundation (NSF) Grant No. PHY-0354401. We thank
Carl H. Albright for communication about original lopsided model.

\appendix

\end{document}